# Dynamic Origin of Evolution and Social Transformation

A.P. Kirilyuk

*G. V. Kurdyumov Institute for Metal Physics,*
*National Academy of Sciences of Ukraine,*
*36 Academician Vernadsky Blvd.,*
*UA-03142 Kyyiv, Ukraine*

We analyse the unreduced, nonperturbative dynamics of an arbitrary many-body interaction process with the help of the generalised effective potential method and reveal the well-specified universal origin of change (emergence), time and evolution in an a priori conservative, time-independent system. It appears together with the universal dynamic complexity definition, where this unified complexity conservation and transformation constitutes the essence of evolution. We then consider the detailed structure of this universal evolutionary process showing its step-wise, "punctuated" character, now provided with the exact mathematical description. Comparing the expected features of a revolutionary complexity transition near a step-like complexity upgrade with the currently observed behaviour of world's social and economic systems, we prove the necessity of complexity revolution towards the superior civilisation level of well-defined nature, the only alternative being an equally dramatic and irreversible degradation, irrespective of efforts applied to stop the crisis at the current totally saturated complexity level.

Ми аналізуємо нередуковану, непертурбативну динаміку довільного процесу взаємодії багатьох тіл за допомогою узагальненого методу ефективного потенціалу і розкриваємо точно визначене універсальне походження зміни (виникнення), часу та еволюції в апріорі консервативній, незалежній від часу системі. Воно з'являється разом з визначенням універсальної динамічної складності, згідно з яким збереження та перетворення цієї уніфікованої складності і створює еволюцію. Далі ми розглядаємо детальну структуру цього універсального еволюційного процесу яка демонструє свій ступінчастий, переривистий характер споряджений тепер точним математичним описом. Порівнюючи очікувані особливості революційного переходу складності біля її ступінчастого підвищення з сучасною поведінкою світових суспільних та економічних систем, ми доводимо необхідність революції складності до вищого рівня цивілізації добре визначеної природи, з єдиною можливою альтернативою такої ж драматичної та необоротної деградації, незалежно від зусиль які вживаються на подолання кризи на сучасному цілком насиченому рівні складності.

Мы анализируем нередуцированную, непертурбативную динамику произвольного процесса взаимодействия многих тел с помощью обобщённого метода эффективного потенциала и выявляем хорошо определённую универсальную природу изме-нения (возникновения), времени и эволюции в априори консервативной, независящей от времени системе. Она обнаруживается вместе с определением универсальной динамической сложности, сохранение и преобразование которой и составляет сущность эволюции. Мы затем рассматриваем детальную структуру этого универсального эволюционного процесса, демонстрирующую его ступенчатый, прерывистый характер, снабжённый теперь точным математическим описанием. Сравнивая ожидаемые особенности революционного перехода сложности вблизи её ступенчатого роста с наблюдаемым сейчас поведением мировых общественных и экономических систем, мы доказываем необходимость революции сложности к высшему уровню цивилизации хорошо определённой природы, с единственной возможной альтернативой столь же драматической и необратимой деградации, независимо от усилий прилагаемых для преодоления кризиса на текущем, полностью насыщенном уровне сложности.

## 1. INTRODUCTION

In our time of great and rapid changes in natural, technical and social systems the origin, direction and efficiency of evolutionary processes is of special, not only theoretical but also increasingly practical interest. Despite progressively growing efforts to put the evolution theory on a firm rigorous basis, the problem – often formulated also in terms of the origin of time – remains practically unsolved within conventional analysis. Two major, fundamentally separated approaches, statistical and dynamical ones, are reduced to mere postulation of empirically observed changes, either in the form of permanently growing entropy (usually for relatively gradual and smoothly distributed changes), or in the form of model-based dynamical structure formation with artificially inserted time variable (for stronger and uneven changes).

      Recent nonperturbative analysis of real many-body interaction problem with arbitrary interaction potential reveals a qualitatively new, totally dynamic origin of change, time and randomness in the form of fundamental dynamic multivaluedness, or redundance, of unreduced interaction results [1-11]. In this paper we review this analysis (section 2) and the ensuing origin of any system evolution (section 3). We then consider its application to social system evolution and transformation, with special attention to current critical development problems (section 4). We thus show how the obtained mathematically rigorous and now truly complete solution to unreduced many-body interaction problem leads to consistent un-

derstanding of modern critical point of human species and civilisation evolution and successful transition to its progressive branch [8]. We conclude with an overview of major features and perspectives of the expected new kind of civilisation after this nontrivial "phase transition" (section 5).

## 2. UNREDUCED MANY-BODY INTERACTION

We analyse arbitrary (pair-wise) interaction in any many-body system with the help of a general Hamiltonian equation for a distribution function called here *existence equation* and coinciding in form with the quantum mechanical Schrödinger equation or classical Hamilton-Jacobi equation [1-11]. We later show (section 3) that it is indeed the universal description of any many-body interaction. The existence equation actually just describes the initial configuration of a system of $N$ interacting entities:

$$\left\{ \sum_{k=0}^{N} \left[ h_k(q_k) + \sum_{l>k}^{N} V_{kl}(q_k, q_l) \right] \right\} \Psi(Q) = E\Psi(Q), \qquad (1)$$

where $h_k(q_k)$ is the generalised Hamiltonian describing the (known) dynamics of the $k$-th system component with its degrees of freedom $q_k$, $V_{kl}(q_k, q_l)$ is the arbitrary interaction potential for the $k$-th and $l$-th components, $\Psi(Q)$ is the system state-function fully describing its configuration $Q \equiv \{q_0, q_1, ..., q_N\}$ and $E$ is the generalised Hamiltonian eigenvalue (generalised energy). As becomes clear in further analysis (section 3), this generalised Hamiltonian/energy represents a universal measure of dynamic complexity defined below (thus extending respective usual notions). Explicit time dependence, if any, enters the same description of eq. (1) by energy replacement on the right with a time derivative operator.

One can conveniently rewrite the general interaction problem formulation of eq. (1) in terms of known eigen-modes of the system components, which gives an equivalent system of equations [1-11]:

$$\left[ h_0(\xi) + V_{nn}(\xi) \right] \psi_n(\xi) + \sum_{n' \neq n} V_{nn'}(\xi) \psi_{n'}(\xi) = \eta_n \psi_n(\xi), \qquad (2)$$

where $\xi \equiv q_0$ is a special, common degree of freedom (usually system component or configuration space coordinates), $V_{nn'}(\xi)$ are matrix elements of interaction potential between component eigen-modes numbered by $n, n'$ and $\eta_n = E - \varepsilon_n$, with eigen-mode eigenvalues $\varepsilon_n$ (see [1-11] for mode details).

As the problem remains nonintegrable for arbitrary interaction potential and more than two system components, usual approach proceeds with its dimensional reduction to a severely simplified but explicitly or approximately integrable "model", such as

$$\left[h_0(\xi) + V_{nn}(\xi)\right]\psi_n(\xi) = \eta_n \psi_n(\xi) \tag{3}$$

for eqs. (2), with any integrable potential $V_{nn}(\xi)$. Thus obtained explicit solution to that another, simplified problem of eq. (3) involves, however, not only significant and irreducible departures from reality but especially fundamental absence of any true novelty, the desired evolutionary change and related *intrinsic* time flow. Trying to find these features in the *unreduced* interaction process while preserving its *analytical* description, one may start with the straightforward substitution of variables in the original system of eqs. (2) formally reducing system dimensionality but at the expense of equivalent, much more complicated and essentially nonlinear effective interaction potential [1-11]. This method known as "optical potential" in the theory of scattering (e.g. [12, 13]) leads to an equation externally resembling a model description of eq. (3),

$$\left[h_0(\xi) + V_{\text{eff}}(\xi;\eta)\right]\psi_0(\xi) = \eta \psi_0(\xi), \tag{4}$$

but where the *effective potential (EP)* $V_{\text{eff}}(\xi;\eta)$ contains the unreduced interaction complexity in the form of its (nonlinear) dependence on the eigenvalues and eigenfunctions to be found:

$$V_{\text{eff}}(\xi;\eta)\psi_0(\xi) = V_{00}(\xi)\psi_0(\xi) + \sum_{n,i} \frac{V_{0n}(\xi)\psi_{ni}^0(\xi)\int_{\Omega_\xi} d\xi' \psi_{ni}^{0*}(\xi')V_{n0}(\xi')\psi_0(\xi')}{\eta - \eta_{ni}^0 - \varepsilon_{n0}}, \tag{5}$$

where $\varepsilon_{n0} = \varepsilon_n - \varepsilon_0$, $\eta \equiv \eta_0$ is the eigenvalue to be found and $\{\psi_{ni}^0(\xi)\}$, $\{\eta_{ni}^0\}$ are the complete sets of (a priori unknown) eigenfunctions and eigenvalues for a system of equations of smaller dimensionality reduced from the full system of eq. (2) [1-11].

Whereas usual applications of the optical potential method proceed with perturbative reduction of this always nonintegrable EP expression, eq. (5), inevitably implying the same fundamental deficiency, the unreduced EP formalism analysis [1-12] reveals indeed a qualitatively new phenomenon of interaction result splitting into many intrinsically complete and therefore *incompatible* system configurations, or *realisations*, just giving rise to intrinsic time flow and dynamic origin of evolutionary changes. This *dynamic multivaluedness*, or *redundance*, phenomenon can be detected by directly counting the number of eigenvalues $\eta$ for the characteristic equation of eq. (4) with the unreduced EP of eq. (5). It results from the nonlinear EP dependence on $\eta$ reflecting the full complexity of interaction feedback loops in a real many-body system. We thus discover that the total number of incompatible system realisations $N_\Re$ is determined by the number of its interacting eigen-modes, $N_\Re = N_\xi$, where $N_\xi$ is the number of terms in summation over $i$ in eq. (5). This algebraic analysis result is totally confirmed and further supported by its geometric version [1,14] clearly demonstrating the eigenvalue distribution dynamics.

The fundamental importance of this new, intrinsic quality of dynamic redundance of *any real* (unreduced) interaction process is that it provides the desired universal origin of physically real, irreversibly flowing time, evolutionary change and (new) structure formation (or "self-organisation"). Indeed, being *dynamically equal* and physically *incompatible*, those multiple system realisations are forced, by the same driving interaction, to permanently and irreversibly *replace each other*, in a *dynamically random* order thus defined. In other words, realisation plurality implies *fundamental dynamic instability* of each individual realisation that follows its physically transparent cycles of emergence, saturation and inevitable replacement by a next emerging, randomly (dynamically) chosen realisation [1-12,14]. In this process the system incessantly repeats the cycles of *dynamic entanglement* of its interacting degrees of freedom (at the realisation formation stage) and their further disentanglement during transition to the next realisation through a special, *intermediate realisation* of the *generalised wavefunction* with transiently quasi-free components [1-11] (it generalises the now causally understood quantum-mechanical wavefunction at the lowest interaction levels to the distribution function at any interaction/complexity level [1,2,11,15]). Universally defined system change in that realisation rotation process corresponds to well-defined leaps of *physically real time* (see section 3) of a given complexity level, which thus *unstoppably flows* simply due to the driving (multivalued!) interaction process and *cannot be reversed* even artificially because of the *dynamically random* choice of each next system realisation.

As a result, the observed system density $\rho(\xi,Q)$ ($=|\Psi(\xi,Q)|^2$ or $\Psi(\xi,Q)$, depending on complexity level) or any other quantity is obtained as a *dynamically probabilistic sum* of this quantity for all realisations im-plying their permanent dynamically random change:

$$\rho(\xi,Q) = |\Psi(\xi,Q)|^2 = \sum_{r=1}^{N_\Re \oplus} \rho_r(\xi,Q) = \sum_{r=1}^{N_\Re \oplus} |\Psi_r(\xi,Q)|^2 , \qquad (6)$$

where $\Psi_r(\xi,Q)$ is the *r*-th regular (non-intermediate) realisation state function obtained from solution of the unreduced EP formalism, eqs. (4)-(5), $\oplus$ sign stands for the dynamically probabilistic sum character, and the *dynamically determined (a priori) probability* value for the *r*-th realisation emergence, $\alpha_r$, is attached:

$$\alpha_r = \frac{1}{N_\Re} , \quad \sum_r \alpha_r = 1 . \qquad (7)$$

It becomes clear why any usual, dynamically single-valued, either statistical or dynamical-model description cannot reveal any intrinsic origin of time and evolutionary change. Although the former, statistical (or stochastic) analysis formally postulates an imitation of random changes (without revealing their dynamic origin), it is forced then to deal only with their averaged description just at the level of distribution function (our intermediate realisation), perfectly reproducing thus the dynamically single-valued reduction scheme. The same is true for all other dynamically single-valued imitations of "complexity" and emergence in usual theory, including "(strange) attractors", "multistability" and "exponentially diverging trajectories" (or "Lyapunov exponents") as they all deal with formally postulated system "evolution" in mathematical "time" within one and the same realisation, with totally compatible structure parts (some of these approaches, in particular attractors, also deal with system evolution in abstract, "phase" spaces, which deforms essentially the meaning of "change"). One can say that any such dynamically single-valued, or *unitary*, description actually deals with a point-like, zero-dimensional *projection* of the unreduced, dynamically "multi-dimensional" (multivalued) evolution of real system, the former reproducing only the respective strongly limited scope of essential properties of the latter.

These conclusions are confirmed by the *universal definition of dynamic complexity* within our dynamically multivalued description of arbitrary interaction process (in any real system). *Dynamic complexity*, *C*, of a system or interaction process (thus any object) is universally defined as a function of the number of its realisations (or rate of their change) equal to zero for the (unrealistic) case of one system realisation [1-11,14]:

$$C = C(N_\Re), \ dC/dN_\Re > 0, \ C(1) = 0, \qquad (8)$$

where, for example, $C(N_\Re) = C_0 \ln(N_\Re)$ or $C(N_\Re) = C_0(N_\Re - 1)$. As for any real system $N_\Re > 1$ and for macroscopic systems the total $N_\Re$ is a huge number, any real system complexity has a positive (and usually relatively great) value. Correspondingly, all unitary, dynamically single-valued models of usual description, including their imitations of "complexity", correspond to strictly zero value of this universal dynamic complexity. The unreduced dynamic complexity, genuine dynamic randomness, essential evolutionary change (emergence) and physically real, irreversibly flowing time come thus all together as unified manifestations of fundamental dynamic multivaluedness of any real interaction process. It implies that system realisations entering the unreduced complexity definition of eq. (8) should not be confused or tacitly substituted with any loosely defined or empirically observed system "states" or "structure elements", but should instead be explicitly derived as those internally complete results of real interaction development, as shown by the above unreduced EP method. On the other hand, for each particular problem one can often ignore realisations of certain lower (e.g. quantum) levels of complexity that provably do not directly influence the higher-level (e.g. classical) dynamics under consideration.

The hierarchical, multilevel structure of world's complexity thus defined is implied already by the basic EP formalism of eqs. (4)-(5). Indeed, the same analysis can be applied to the reduced system of equations giving rise to eigen-solutions $\{\psi_{ni}^0(\xi)\}, \{\eta_{ni}^0\}$, leading to their dynamical splitting by the same mechanism and so on for a series of all lower-dimension solutions. As a result, one obtains the causally complete final solution and system structure in the form of *dynamically probabilistic fractal*, containing many levels of *permanently randomly changing* realisations of progressively decreasing scale, which gives rise to the rigorous and universal definition of *life* [1,2,6,7,10]. It extends essentially the simplified, abstract and dynamically single-valued model of ordinary fractals, with their limited, always practically broken scale symmetry being now replaced by the externally irregular but always exact *symmetry of complexity* underlying thus real evolution dynamics (see section 3) [1,2,5-11,15].

This most complete, dynamically fractal general solution to the starting unreduced interaction problem of eq. (1) can be presented as the multilevel extension of one-level version of eqs. (6), (7):

$$\rho(\xi, Q) = \sum_{r,r',r''\ldots}^{N_\Re} {}^\oplus \rho_{rr'r''\ldots}(\xi, Q) , \qquad (9)$$

where $r, r', r''...$ enumerate respective fractal level realisations, while the dynamically determined probability $\alpha_{rr'r''...}$ of realisation emergence is

$$\alpha_{rr'r''...} = \frac{N_{rr'r''...}}{N_{\Re}} \ , \qquad \sum_{r,r',r''...} \alpha_{rr'r''...} = 1 \ , \qquad (10)$$

with $N_{rr'r''...}$ being the number of empirically inseparable elementary realisations within the corresponding observed composite realisation. The permanent "horizontal" realisation change at any level is completed here by "vertical" structure development to other levels, which has the transparent physical interpretation of progressive emergence of new structures as a result of interaction of structures formed at neighbouring levels. This "evident" interaction process development and real structure creation would be impossible, however, without much less evident dynamic multivaluedness at each interaction level. Even the average expectation value $\rho_{\text{ex}}(\xi, Q)$ (for long enough observation time) hides in it a very complicated, multivalued and multilevel interaction development process:

$$\rho_{\text{ex}}(\xi, Q) = \sum_{r,r',r''...}^{N_{\Re}} \alpha_{rr'r''...} \rho_{rr'r''...}(\xi, Q) \ . \qquad (11)$$

## 3. SPACE, TIME, EVOLUTION AND THE UNIVERSAL SYMMETRY OF COMPLEXITY

We can now provide the above physical origin of time (emergence) and evolution (structure formation) in the multivalued interaction dynamics (section 2) with a rigorous expression. We first note that what actually emerges in such real, dynamically multivalued interaction process is different, permanently mutually replaced system realisations forming its evolving structure. Therefore one can start with universal definition of *elementary space (structure) element*, $\Delta x$, as characteristic eigenvalue separation for the unreduced EP formalism, eqs. (4)-(5), $\Delta x = \Delta \eta_i^r$, where $r$ enumerates realisations and $i$ eigenvalues within the same realisation. One should distinguish here between the *elementary length* of system jump between realisations, $\lambda = \Delta x_r = \Delta_r \eta_i^r$ (neighbouring $r$ values, fixed $i$), determining the dimension of observed dynamic structures, and the *minimum size* of effective *space point*, $r_0 = \Delta x_i = \Delta_i \eta_i^r$ (fixed $r$, neighbouring $i$ values), reflecting the smallest system dimension at a given complexity level. Based exclusively on the unreduced (multivalued) interaction develop-ment, we obtain thus the totally consistent and universal definition of *intrinsically discrete* and *physically tangible* space structure resulting from that interaction.

As physically real time (and evolution) originates from system realisation change (section 2), one obtains now *universal time definition* as intensity specified as frequency $\nu$ of realisation change, with *elementary time interval (period)* $\Delta t$ for that frequency being $\Delta t = \tau = 1/\nu = \lambda/\upsilon_0$, where $\lambda = \Delta x_r$ is the above elementary length of system jump between realisations and $\upsilon_0$ is the signal propagation speed in the material of interacting components (known from lower complexity levels). Thus defined real time is *permanently flowing* due to *unstoppable* transitions between system realisations driven by its interaction and this time flow is *irreversible* because of the *dynamically random* choice of each next realisation.

Note that in usual, dynamically single-valued interaction models there is only one realisation and therefore $\lambda = \Delta x_r = 0$, $\Delta t = \lambda/\upsilon_0 = 0$, so that there can be no either genuine structure formation or real time flow, both of them being inserted only artificially (postulated), including conventional "self-organisation" and "chaos" theories. By contrast, in our unreduced, dynamically multivalued description there is a multilevel, *fractally structured hierarchy* of real space and time corresponding to the hierarchy of developing interaction complexity [1,5,8,10,11,15]. Whereas each individual (big enough) level of complex dynamics is observed and characterised as stationary system *mechanics*, transitions between essentially different complexity levels appear as explicit *evolution* phenomena, even though in both cases one deals with the same process of dynamically multivalued interaction development. We shall specify now both these cases within a unified description in terms of suitable complexity measures.

As the elementary space and time intervals introduced above are determined by system transitions between realisations, a major physical measure of complexity determined by the number of realisations (eq. (8)) emerges as the generalised *action* $\mathcal{A}$ as the simplest quantity proportional to both time and space (increments) and now extended to any level of complex world dynamics [1,5,8,10,11]:

$$\Delta \mathcal{A} = p\Delta x - E\Delta t, \quad (12)$$

with coefficients $p$ and $E$ recognised as (now generalised) *momentum* and (total) *energy*:

$$p = \frac{\Delta \mathcal{A}}{\Delta x}\Big|_{t=\text{const}} \simeq \frac{\mathcal{A}_0}{\lambda}, \quad (13)$$

$$E = -\frac{\Delta \mathcal{A}}{\Delta t}\bigg|_{x=\text{const}} \simeq \frac{\mathcal{A}_0}{\tau} \ , \qquad (14)$$

where $\mathcal{A}_0$ is a characteristic action value at a given complexity level, while $x$ and $p$ should be properly understood as vectors where necessary. We see thus that these omnipresent quantities, momentum and energy, are also universal *differential measures of complexity*, whereas action is its *integral* measure, which extends essentially the meaning and importance of these originally mechanical quantities. We can see also that, following space and time discreteness, action is a *dynamically discrete*, or naturally *quantised*, quantity at any level of complexity, leading to fundamental quantum-mechanical discreteness at the lowest complexity levels (with the elementary action increment $\mathcal{A}_0 = \hbar$, Planck's constant) [1,2,11,15] but also to discrete structures and evolutionary transitions at higher complexity levels (section 4).

For irreversibly growing time and always positive total energy, complexity-action $\mathcal{A}$ as determined by eq. (12) will always decrease with time, irrespective of interaction development details. It is but another expression of dynamically random realisation choice underlying time irreversibility [1,5,8,10,11,15]. On the other hand, there is certainly also a growing form of complexity in the same (arbitrary) interaction process development that describes emergence of ever growing number of its fractally structured realisations (in agreement with the universal complexity definition of eq. (8)). It generalises traditional entropy to any (real) interaction process, and therefore we call this growing complexity form *dynamic entropy*, *S*. Because both decreasing complexity-action and growing complexity-entropy account for one and the same process of interaction-driven structure emergence, with the same underlying universal definition of dynamic complexity of eq. (8), it becomes clear that one of them, complexity-entropy, grows exactly at the expense of the dual form of complexity-action, so that the decrease of action, also called *dynamic information I* [1,2,5,7-11,15], produces just the same quantity of dynamic entropy. Whereas dynamic information expressed as action accounts for system (interaction) *potentiality* to produce new structures (realising thus the universal integral extension of "potential energy"), dynamic entropy describes the irreversibly produced *tangible result* of that potential power, in the form of real-structure complexity.

In summary, any interaction process can be universally described as *conservation and transformation*, or *symmetry*, *of the total dynamic* *complexity C* defined as the *sum of dynamic information (action)* $I = \mathcal{A}$ *and dynamic entropy S* (measured in the same units of generalised action), $C = I + S = \mathcal{A} + S$, where the first summand permanently decreases to the exact amount of simultaneous growth of the second summand:

$$\Delta C = \Delta \mathcal{A} + \Delta S = 0, \quad \Delta S = -\Delta \mathcal{A} > 0. \qquad (15)$$

Dynamic complexity-entropy of real emerging structures thus simply realises their "plan" described by dynamic complexity-action, while the entire process is a result of the *exact symmetry* (conservation) of total complexity, rather than any conventional extremum principle (e.g. maximum entropy or often evoked maximum or minimum entropy growth rate, etc.). Symmetry of complexity is derived thus as the absolutely *universal* law eventually underlying *all* (correct) particular laws and "principles" always only empirically postulated in usual theory [1,5,8-11,15] (see also below). Contrary to regular, always limited and somewhere broken symmetries of unitary theory, the universal symmetry of complexity remains always exact but relating externally irregular structures and sequences. As it simply connects system's consecutive dynamical states, it also represents the most precise expression of any system's *evolution* understood now as the *necessary* result of dynamic *conservation* by *inevitable* transformation of total *dynamic complexity*. Internally irregular and chaotic *change* is seen now *as a result* of a *perfect and universal symmetry*, contrary to opposite ideas of unitary (dynamically single-valued) science, where change is rather a (conceivably small) deviation from (always inexact) particular symmetry due to its explicit violation by an extraneous influence.

If we consider now manifestations of the universal symmetry of complexity for the case of system *mechanics* (see above), i.e. relatively small changes at a given complexity level, then we can produce its more convenient, differential form by dividing eq. (15) by a small *real time* increment $\Delta t|_{x=\text{const}}$ (consistently defined above):

$$\frac{\Delta \mathcal{A}}{\Delta t}\Big|_{x=\text{const}} + H\left(x, \frac{\Delta \mathcal{A}}{\Delta x}\Big|_{t=\text{const}}, t\right) = 0 \ , \quad H = E > 0 \ , \qquad (16)$$

where the *generalised Hamiltonian*, $H = H(x, p, t)$, is the differential expression of complexity-entropy, $H = (\Delta S/\Delta t)|_{x=\text{const}}$, in agreement with the definition of generalised (total) energy $E\ (=H)$ through complexity-action, eq. (14), and generalised momentum definition, eq. (13). We obtain thus the discrete complex-dynamic extension of the well-known Hamilton-Jacobi equation provided now with a new, essentially generalised meaning and time-flow direction towards growing dynamic entropy and decreasing dynamic information (complexity-action). The latter condition can be further amplified, if we introduce the *generalised Lagrangian*, $L$, as the (generally discrete) total time derivative of complexity-action:

$$L = \frac{\Delta \mathcal{A}}{\Delta t} = \frac{\Delta \mathcal{A}}{\Delta t}\Big|_{x=\text{const}} + \frac{\Delta \mathcal{A}}{\Delta x}\Big|_{t=\text{const}} \frac{\Delta x}{\Delta t} = pv - E = pv - H , \quad (17)$$

where $v = \Delta x/\Delta t$ is the (global) motion speed and the scalar product of vectors is implied if necessary. The same fundamental feature of dynamically random choice among multiple incompatible realisations implies permanently decreasing dynamic information of action:

$$L < 0, \quad H, E > pv \geq 0 , \quad (18)$$

which is the generalised and stronger version of the extended second law of eqs. (15), (16) determining the time arrow direction.

The *generalised Hamilton-Jacobi formalism* of eq. (16) describing the evolution of "regular", localised and entangled, system realisations can be completed with the equally universal *Schrödinger equation* for the *generalised wavefunction, or distribution function, $\Psi$* of intermediate, delocalised and disentangled, realisation (section 2) [1,5,8-11]:

$$\mathcal{A}_0 \frac{\Delta \Psi}{\Delta t}\Big|_{x=\text{const}} = \hat{H}\left(x, \frac{\Delta}{\Delta x}\Big|_{t=\text{const}}, t\right) \Psi(x,t) , \quad (19)$$

where $\mathcal{A}_0$ is a characteristic action value from the generalised quantisation rule (see below) that may include a numerical constant ($\mathcal{A}_0 = i\hbar$ for quantum complexity levels), while the Hamiltonian operator, $\hat{H}(x,\hat{p},t)$, is obtained from its ordinary form of eq. (16) by replacement of momentum variable $p = (\Delta \mathcal{A} / \Delta x)|_{t=\text{const}}$ with the respective "momentum operator", $\hat{p} = \mathcal{A}_0 (\Delta / \Delta x)|_{t=\text{const}}$. The generalised Schrödinger equation, eq. (19), is related to the generalised Hamilton-Jacobi equation, eq. (16), by the *dynamic quantisation rule*,

$$\Delta(\mathcal{A}\Psi) = 0, \quad \Delta \mathcal{A} = -\mathcal{A}_0 \frac{\Delta \Psi}{\Psi} , \quad (20)$$

which results from the same dynamic complexity conservation law of eq. (15) (first equality) implying here that each regular realisation is obtained by intermediate realisation (generalised wavefunction) "reduction" due to entanglement of interaction components, with further disentanglement back to intermediate realisation [1,5,8-11,15]. The same multivalued dy-namics for the case of measurement process leads to the *generalised Born's rule* [1,5,8-11,15] providing another, extremely convenient expression for regular realisation probabilities (cf. eqs. (7), (10)):

$$\alpha_r = |\Psi(X_r)|^2 \ , \tag{21}$$

where $X_r$ is the *r*-th realisation configuration and $\alpha_r$ its probability. This extension of respective relation of usual quantum mechanics (simply postulated but never causally derived there) to arbitrary system dynamics explains the importance of the generalised Schrödinger equation, especially for cases of sufficiently "smeared" dynamics with many close realisations.

The resulting *Hamilton-Schrödinger formalism*, eqs. (16)-(21), expresses thus the universal symmetry of complexity, eq. (15), especially for cases of relatively smooth system evolution within the same (big) complexity level. We can see now the origin of universality of the starting Hamiltonian description, eq. (1), referred to at the beginning of section 2. Moreover, we can see how the symmetry of complexity and its Hamilton-Schrödinger formalism underlies (and now unifies) many popular, actually postulated dynamic equations and principles. For example, if we consider generalised Hamiltonian expansion in powers of its momentum variable,

$$H(x,p,t) = \sum_{n=0}^{\infty} h_n(x,t) p^n \ , \tag{22}$$

with generally arbitrary functions $h_n(x,t)$, then its substitution into the generalised Hamilton-Jacobi and Schrödinger equations gives respectively (for ordinary, continuous versions of derivatives):

$$\frac{\partial \mathcal{A}}{\partial t} + \sum_{n=0}^{\infty} h_n(x,t) \left(\frac{\partial \mathcal{A}}{\partial x}\right)^n = 0 \ , \tag{23}$$

$$\mathcal{A}_0 \frac{\partial \Psi}{\partial t} = \sum_{n=0}^{\infty} h_n(x,t) \frac{\partial^n \Psi}{\partial x^n} \ . \tag{24}$$

For various $h_n(x,t)$ and series truncations one can obtain here many particular model equations. Other models result from simplification of dynamically nonlinear dependence of unreduced EP, eqs. (4)-(5), on eigenfunctions and eigenvalues to be found. In addition to universal conservation and transformation of complexity (including generalised first and second laws of thermodynamics), one can derive other fundamental laws and principles from this universal description of multivalued dynamics, including now causally substantiated quantum behaviour and elementary particle properties intrinsically unified with equally dynamically explained laws of special and general relativity [1,11,15].

If we consider now the situation of very uneven, "revolutionary" transformation of complexity-action to complexity-entropy *between big enough complexity levels* (complementary to the above *mechanics* within each level), then the universal Hamilton-Schrödinger formalism of eqs. (16)-(21) will be much less useful because of "singular", relatively great complexity (and structure formation) leap involved. One may analyse such transitions in more detail applying other, more qualitative approach to manifestations of the same symmetry of complexity of eq. (15).

A general scheme of evolutionary complexity transformation process is shown in Fig. 1 [8]. Here the characteristic increment, $\Delta S$, of dynamic complexity-entropy during system "revolutionary" transition from *i*-th to *j*-th complexity level is much greater than its maximum variations, $\Delta S_i, \Delta S_j$, in each level dynamics, $\Delta S \gg \Delta S_i, \Delta S_j \sim \mathcal{A}_{0i}, \mathcal{A}_{0j}$, where $\mathcal{A}_{0i}, \mathcal{A}_{0j}$ are characteristic (absolute) action values for respective levels. Therefore complexity evolution analysis in terms of differential equations, eqs. (16), (19), becomes inefficient on this global scale. However, we can clearly specify the fundamental origin of both system evolution as such and its strongly uneven, step-wise dynamics clearly seen in Fig. 1. It is reduced to dynamic multivaluedness of any real interaction process (section 2) giving rise to permanent, irreversible (dynamically random) realisation change and new structure formation, as well as inevitable realisation discreteness (related to dynamic entanglement of interacting system components) taking relatively huge proportions at those greater, revolutionary transitions (involving proportionally greater system volumes and dynamical depth).

Note once again the related important feature of permanent dynamic entropy *growth* during evolutionary process in our *dynamically multivalued* description (in full agreement with the generalised second law of thermodynamics), as opposed to the well-known persisting problem in usual dynamically single-valued theory, where the appearance of new externally ordered structures enters in contradiction with the second law (usually "solved" by incorrect reference to system openness to a greater environment becoming unnecessary in our dynamically multivalued description). We obtain thus the totally correct (and strong) evolution possibility (and even necessity) in a totally isolated system, as any new, however externally "ordered" structure contains a huge multiplicity of new, cha-otically changing realisations and corresponds thus to essential entropy growth [1,2,4,5,7-11].

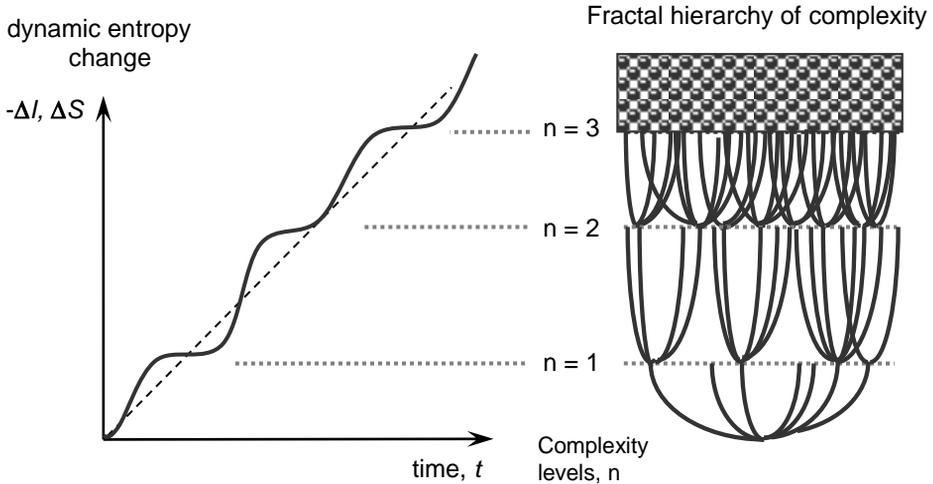

**Fig. 1.** Scheme of universal system evolution by permanent transformation of its dynamic complexity-information (*I*) into complexity-entropy (*S*).

Note that the mentioned and other evident, persisting (and pressing!) contradictions of usual, dynamically single-valued theory lead to recognition of fundamental failure of respective traditional, positivistic science method, with striking conclusions and far-reaching beliefs, such as the alleged absence of any scientifically certain law governing arbitrary, complex enough system evolution [16]. As we can clearly see now, this is only an artefact of conventional, positivistic science paradigm that replaces the (absent) correct solution of unreduced many-body interaction problem with its incorrect "model" approximations neglecting all but one system realisations. By contrast, the unreduced, dynamically multivalued problem solution (section 2) naturally leads to general evolution law, eq. (15), that can be specified down to respective particular laws, eqs. (16)-(21), or even (now correctly derived) dynamic models that should always be analysed, however, within a dynamically multivalued description, such as the universal EP method applied above. The obtained universal evolution law and its versions include dynamically specified (true) randomness, uncertainty/probability and irreversibility as its integral constituents.

## 4. THE STRUCTURE OF COMPLEXITY TRANSITION AND SOCIAL TRANSFORMATION

Let us now consider the detailed structure of such greater complexity transition illustrated in Fig. 2, with explicit reference to a social system transformation [8] (although the analysis remains, of course, absolutely universal and applicable to any system evolution).

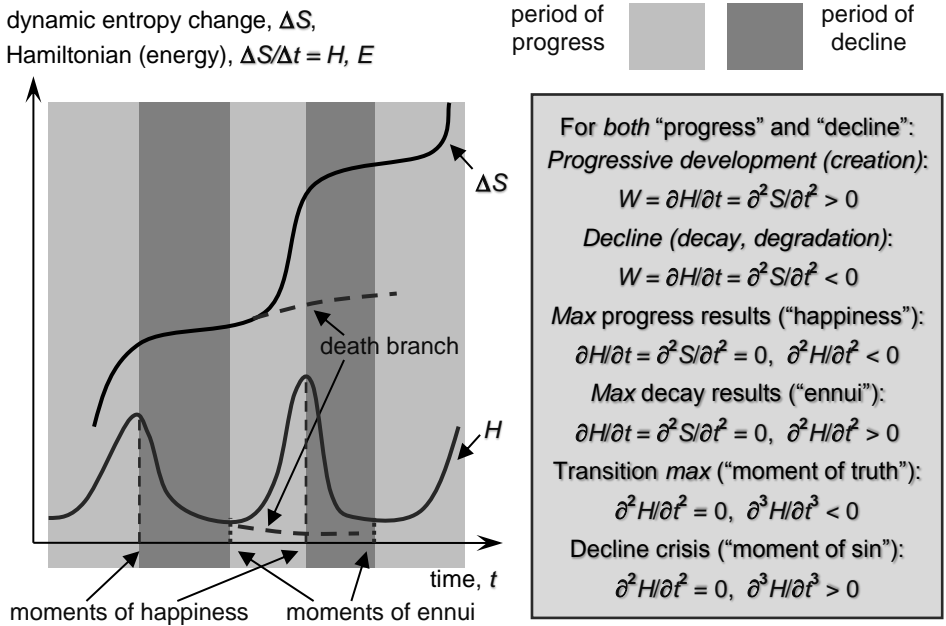

**Fig. 2.** Universal periods of a real system progress, decline and transitions between them in terms of dynamic complexity-entropy (or action) change $\Delta S = -\Delta \mathcal{A}$, generalised Hamiltonian $H = \partial S/\partial t$ or energy $E = -\partial \mathcal{A}/\partial t = H$ and higher derivatives.

　　　This graphical illustration shows a part of step-like complexity-entropy growth (evolutionary) curve of Fig. 1 completed with respective time dependence of its (partial) time derivative, the Hamiltonian $H = \partial S/\partial t$ or energy $E = -\partial \mathcal{A}/\partial t = H$. As shown in the insert on the right, higher time derivatives of dynamic complexity-entropy also play important role in practical tendency analysis marking various critical (and easily recognisable) moments in social system evolution around its big complexity transition [8]. It should be emphasized that the underlying definition of dynamic complexity, eq. (8), takes into account all system inter-actions, in our case all social interactions in all their real economic, political, psychological and other aspects, rather than their various severely limited "models" emphasizing separate (e.g. economical) aspects. The resulting social complexity development laws illustrated in Fig. 2 and having clearly observable, strong manifestations possess therefore totally objective, (new) exact-science nature liberated from usual social-science uncertainties, which implies complete reliability of conclusions derived from those observable manifestations.

A major feature of this universal complexity development curve in Fig. 2 is the *complexity transition*, or *revolution*, which is a sharp and high step-wise rise on the temporal dependence of dynamic entropy-complexity ( $\Delta S$ ) corresponding to a narrow maximum on the temporal dependence of generalised energy ( $E = H = \partial S/\partial t$ ) and separating neighbouring periods of progress and decline. If we start at the end of the last decline period, at the "moment of ennui" where $\partial H/\partial t = \partial^2 S/\partial t^2 = 0$, $\partial^2 H/\partial t^2 > 0$, then we enter in a very short *period of bifurcation* of system's dynamic selection between creative (rising) tendency of complexity transition and destructive tendency of the "death branch", which is simply the default continuation of the previous decline tendency in the absence of complexity revolution.

Complexity revolution leading to the superior complexity level occurs if the system has a high enough potential (stock) of hidden interaction complexity-action, or dynamic information, which is to be transformed into explicit complexity-entropy of the emerging higher level of system structure. In the opposite case, for example in the case of too "old" and "tired" social system, there is no enough (potential) energy in the system to perform that big structure transformation and it is condemned to a rapid degradation of the death branch. In this connection we emphasize again the highly uneven, discrete, or "nonlinear" character of this specific phase of complexity transition contrasting with previous smooth evolution, where the former contrary to the latter cannot proceed in a gradual regime of "small steps". That "sudden" (and practically often "unexpected") switch to a qualitatively different regime of change is deeply rooted in the unreduced, holistic interaction dynamics, where the entirely formed system structure of existing levels (attained precisely at that time and not before) constitutes itself the main obstacle for its further "smooth" development. It is related to the physical origin of system realisation discreteness (e.g. quantum-mechanical discreteness at the lowest complexity levels [1-3,11,15]), where a system can only "jump" to another realisation or higher complexity level as a whole, through its complete restructuring (disentan-glement and new entanglement of interacting components) occurring necessarily in a step-wise manner (sections 2, 3). That is why the necessity, origin and dynamics of complexity transition cannot be adequately described within usual, dynamically single-valued theory framework, irrespective of its model sophistication (including computer simulations). Note also the equally important implication of the symmetry (transformative conservation) of interaction complexity that constitutes the underlying integral, genuine reason of system (complexity) development, as opposed to any particular tendency (including falsely understood entropy growth in the unitary theory).

Based on that universal complexity transition dynamics and currently observed economic, social, psychological and bio-ecological tendencies (cf. Fig. 2) we can state therefore that the entire planetary human civilisation – acquiring right now the characteristically unified, "globalised" structure of a "phase transition" – is situated just at that critical bifurcation point of selection between the "revolution of complexity" (transition to the superior level of its dynamic complexity) and the "death branch" of mere irreversible destruction (inevitably ending at a much lower complexity level) [8-10,17]. The unprecedented and actually historically unique scale of (very rapid) divergence between those two incompatible (and the only real) possibilities certainly necessitates equally unprecedented efforts in order to realise the progressive development tendency and avoid the only alternative of self-destruction, the more so that the latter corresponds to the default, "inherited" tendency of previous "natural" (not any more!) smooth growth. Those extraordinary efforts can only be based on the unreduced understanding of complexity transition and its manifestations (see below) within the holistic description of all real social-system interactions uniquely provided by our analysis.

Let us emphasize once more two key, practically important results of this causally complete description of modern critical state of global civilisation development. The first is the fact of unique, unavoidable choice between two qualitatively big changes, those of global progress by complexity revolution and equally rapid degradation within the dominating (default) death branch. Contrary to various, especially economic "models" of dynamically single-valued imitation of real interaction processes, there is no other possibility somewhere "in between" those two extreme, quickly diverging choices. In particular, one cannot separate, especially near this critical point, any particular, e.g. economical aspects of development from other, equally important (e.g. "human") dimensions. Therefore after the complete (unprecedented, including technological) saturation of the current complexity level the system cannot simply find its way out of today's "economic crisis" by analogy to previous economic difficulties occurring within yet *unsaturated* complexity level development.

The second particularly important result uniquely provided by the present analysis is that the necessary progressive change cannot be smaller than the qualitatively big growth of unreduced complexity of civilisation dynamics up to its superior level, which implies a qualitative change of the entire social system structure, including its "human" (intellectual and spiritual) dimensions. This feature strongly limits the scope of suitable changes and provides the indispensable general direction of their realisation. Thus all partial, "technical" system modifications at existing complexity level become now fundamentally, qualitatively insufficient, irrespective of efforts applied (including any resource/effort redistribution and amplification of particular development aspects, such as "education", "computerisation", or "ecology", often evoked as the necessary "revolutionary" change within unitary development concepts). By contrast, based on our unreduced interaction complexity understanding we can specify changes objectively necessary and sufficient in order to realise the revolution of civilisation complexity towards its superior levels.

## 5. COMPLEXITY REVOLUTION AND THE NEXT-LEVEL CIVILISATION

Referring to our more detailed description [8], we can specify now essential features of the superior level of social system complexity determining also the direction of complexity transition towards that next-level civilisation. As various ideas of a necessary social transformation become increasingly popular in this very special epoch of change and uncertainty, one should emphasize first of all what the sustainable new level of complexity cannot be, in order to avoid easy but misleading imitations of that important transition. Namely, one should exclude from consideration any modification of existing, *Unitary System* of social organisation, with its centralised and obligatory (linear) power dynamics and eventually equally linear economic and human relations. The latter may seem to possess greater freedom and complexity than political system as such, at least within any basically liberal version of Unitary System ("market economy", "democracy", etc.). However, eventually every aspect and dimension of such social system dynamics is forced to follow the same unitary, cen-tralised and characteristically limited (artificially and mechanically "enforced") dynamics inevitably ending up in self-destructive saturation.

We can see that this mechanically fixed system of orders cannot overcome its fundamental complexity limits and becomes saturated and obsolete just at the stage of its highest possible perfection corresponding to all tried and imaginable technological applications. Thus today's "financial" problems are not due to limited power of available computer calculations and they can hardly be solved by any increase or "stronger" application of that power. By contrast, in terms of a popular biological analogy of social system, one can say that what is definitely missing is social system's (conscious) *intelligence* considered as a property of the entire social organism (starting, of course, from any national or even smaller scale). It is easy to see indeed that any most allegedly "advanced" version of unitary social system, with all its "scientific" and "intellectual" departments, still represents nothing more than a version of the same primitive tribal organisation, with the eventually dominating "power of the crowd" devoid of any real (collective) mind by definition and only formally delegated to and realised by respective "central units". The limits of complexity development of *any* such unitary system are simply due to its highest possible *empirically* driven use of available resources that inevitably attains (right now) its evident technical limitation due to physically complete (empirical) comprehension and quantitatively limited stock. At that point any version of Unitary System loses any further (general) purpose and thus meaning of existence and becomes inevitably unstable against dissociative degradation.

The only possible alternative to resulting Unitary System destruction and the unique way of further progress can therefore be attained at a superior level of *social conscious intelligence*, or genuine "social mind", with its respective social and individual realisation. This fundamental conclusion is in perfect agreement with our description of (any) consciousness as a high enough (and well-specified) level of the same unreduced dynamic complexity [18]. It is thus the right moment now for any real social organism to acquire this higher level of conscious dynamics or, in other words, to become a truly conscious adult organism, after previous stages of social "childhood" with essentially limited consciousness and basically only empirically driven, animal intelligence. As with any kind of conscious behaviour, it practically implies the prerequisite genuine, *causally complete understanding* of any real situation and way of development or problem solution, here at the level of entire society, which is driven thus by such *power of ideas*, rather than unitary power of individuals (or practical needs). Naturally, this essentially new quality implies serious social structure change and progress towards the one explicitly guided by respective (new) organs and priorities.

This superior level of social structure and thus human civilisation development, provided with the ensuing solutions of known major problems of the degrading Unitary System, can be called the *Harmonical System*, in agreement with its intrinsic sustainability [8]. The superior possibilities of Harmonical System are well illustrated by the phenomenon of *exponentially huge power* of unreduced complex dynamics with respect to any unitary model, related to the dynamically multivalued fractal structure of the former (section 2) [2,3,7-11,17,18]. This enormous, practically "magic" efficiency jump is the unique way to span the current equally impressive and always growing gap between practical development needs and failing unitary system stagnation. We can only mention here major aspects of qualitatively new, harmonical social organisation and dynamics after the jump (all of them rigorously substantiated by progressive complexity growth criterion, cf. section 4), including *emerging* (rather than fixed) *decision power and social structure*, *complexity-increasing production* ways, *new kind of settlement and infrastructure* and the underlying new kind of understanding (and organisation of science) of the *universal science of complexity* (with its unreduced, multivalued dynamics) [8,19]. The latter inevitably becomes thus an *integral (and major) part* of this true *knowledge-based society*, contrary to now dominating but strongly limited unitary, dynamically single-valued science "models" fundamentally separated from any real system dynamics and its consistent understanding, as well as from any technologically "advanced" society dynamics and government. Essential knowledge development from unitary imitative models to causally complete understanding of unreduced, dynamically multivalued real-system complexity is therefore inseparable from, and thus can only occur together with, the necessary social system progress from its ending unitary to the forthcoming harmonical level.

The emerging new civilisation of harmonical level automatically overcomes the tragic destructive purposelessness of the ending unitary civilisation and acquires the universal superior Purpose of now unlimited and dominating progressive growth of complexity-entropy (guided by its superior conscious levels), which corresponds to vanishing depressions on the $H(t)$ curve in Fig. 2. Contrary to old and new unitary religious and ideological imitations, the unified Purpose of harmonical levels is natural-ly integrated into any practical activity, so that there is no more contradiction between the end and the means and no blind domination of the latter. In particular, the Purpose is indistinguishable from the clearly specified "entailing law" (cf. ref. [16]) of the *universal symmetry of complexity*, provided with all necessary emergence and uncertainly aspects of exact dynamic origin (sections 2, 3).

Note finally that as any kind of higher-level "phase transition", this social revolution of complexity can occur as a locally great but spatially gradual, self-propagating process of qualitative change, which essentially simplifies its practical realisation. By contrast, it cannot occur in a locally gradual way implied by any unitary development concept inevitably related to the dominating Unitary System and its way of thinking, which just determines today's critically high evolution barrier and the necessity of complexity revolution in both knowledge and society development.

## REFERENCES

1. A.P. Kirilyuk, *Universal Concept of Complexity by the Dynamic Redundance Paradigm: Causal Randomness, Complete Wave Mechanics, and the Ultimate Unification of Knowledge* (Kyiv: Naukova Dumka: 1997). For a non-technical review see also ArXiv:physics/9806002.
2. A.P. Kirilyuk, "Dynamically Multivalued, Not Unitary or Stochastic, Operation of Real Quantum, Classical and Hybrid Micro-Machines", *ArXiv:physics/0211071*.
3. A.P. Kirilyuk, *Nanosystems, Nanomaterials, Nanotechnologies*, **2**, Iss. 3: 1085 (2004). ArXiv:physics/0412097.
4. A.P. Kirilyuk, *Solid State Phenomena*, **97-98**: 21 (2004). ArXiv:physics/0405063.
5. A.P. Kirilyuk, *Proceedings of Institute of Mathematics of NAS of Ukraine*, **50**, Part 2: 821 (2004). ArXiv:physics/0404006.
6. A.P. Kirilyuk, In: G.A. Losa, D. Merlini, T.F. Nonnenmacher, and E.R. Weibel (Eds.), *Fractals in Biology and Medicine, Vol. III* (Basel: Birkhäuser: 2002), p. 271. ArXiv:physics/0305119.
7. A.P. Kirilyuk, In: G.A. Losa, D. Merlini, T.F. Nonnenmacher, and E.R. Weibel (Eds.), *Fractals in Biology and Medicine, Vol. IV* (Basel: Birkhäuser: 2005), p. 233. ArXiv:physics/0502133; hal-00004330.
8. A.P. Kirilyuk, In: V. Burdyuzha (Ed.), *The Future of Life and the Future of Our Civilisation, Vol. IV* (Dordrecht: Springer: 2006), p. 411. ArXiv:physics/0509234; hal-00008993.
9. A.P. Kirilyuk, In: D. Gaïti (Ed.), *Network Control and Engineering for QoS, Security, and Mobility, IV*, *IFIP, Vol. 229* (Boston: Springer: 2007), p. 1. ArXiv: physics/0603132; hal-00020771.
10. A.P. Kirilyuk, "Universal Science of Complexity: Consistent Understanding of Ecological, Living and Intelligent System Dynamics", *ArXiv:0706.3219*; *hal-00156368*.
11. A.P. Kirilyuk. "Complex-Dynamical Solution to Many-Body Interaction Problem and Its Applications in Fundamental Physics", *ArXiv:1204.3460*; *hal-00687132*.

12. A.P. Kirilyuk, *Nucl. Instr. and Meth.*, **B69**, Iss. 2-3: 200 (1992).
13. P.H. Dederichs, In: H. Ehrenreich, F. Seitz and D. Turnbull (Eds.), *Solid state physics: Advances in research and applications, vol 27* (New York: Academic Press: 1972), p. 136.
14. A.P. Kirilyuk, *Annales de la Fondation Louis de Broglie, 21*, Iss.4: *455 (1996)*. ArXiv:quant-ph/9511034, quant-ph/9511035, quant-ph/9511036.
15. A.P. Kirilyuk, "Consistent Cosmology, Dynamic Relativity and Causal Quantum Mechanics as Unified Manifestations of the Symmetry of Complexity", *ArXiv: physics/0601140*; *hal-00017268*.
16. G. Longo, M. Montévil and S. Kauffman, "No entailing laws, but enablement in the evolution of the biosphere", *ArXiv:1201.2069*.
17. A. Kirilyuk and M. Ulieru, In: M. Ulieru, P. Palensky and R. Doursat (Eds.), *IT Revolutions* (Berlin Heidelberg: Springer: 2009), p. 1. ArXiv:0910.5495; hal-00426699.
18. A.P. Kirilyuk, "Emerging Consciousness as a Result of Complex-Dynamical Interaction Process", *ArXiv:physics/0409140*.
19. A.P. Kirilyuk, In: M. López Corredoira and C. Castro Perelman (Eds.), *Against the Tide: A Critical Review by Scientists of How Physics & Astronomy Get Done* (Boca Raton: Universal Publishers: 2008), p. 179. ArXiv:0705.4562.